\documentclass[10pt,a4paper,aps,showpacs,showkeys,amsmath,amssymb,pra,floatfix,
reprint,superscriptaddress]{revtex4-1}

\usepackage[mathscr]{euscript}

\usepackage{hyperref}

\usepackage{amsmath}
\usepackage{amssymb}
\usepackage{slashed}
\usepackage{xcolor}
\usepackage{graphicx}
\usepackage{grffile}
\usepackage[squaren]{SIunits}
\renewcommand{\mathbf}{\boldsymbol}

\DeclareSymbolFont{rsfs}{U}{rsfs}{m}{n}
\DeclareSymbolFontAlphabet{\mathrsfs}{rsfs}

\renewcommand{\d}{\mathrm d}
\newcommand{\FWHM}{\mathrm{FWHM}}

\newcommand{\resub}[1]{#1}

\begin{document}

\title{Narrowband inverse Compton scattering x-ray sources at high laser intensities}

\author{D.~Seipt}
\email{d.seipt@gsi.de}
\affiliation{Helmholtz-Institut Jena, Fr{\"o}belstieg 3, 07743 Jena, Germany}

\author{S.~G.~Rykovanov}
\affiliation{Helmholtz-Institut Jena, Fr{\"o}belstieg 3, 07743 Jena, Germany}

\author{A.~Surzhykov}
\affiliation{Helmholtz-Institut Jena, Fr{\"o}belstieg 3, 07743 Jena, Germany}

\author{S.~Fritzsche}
\affiliation{Helmholtz-Institut Jena, Fr{\"o}belstieg 3, 07743 Jena, Germany}
\affiliation{Universit\"at Jena, Institut f\"ur Theoretische Physik, 07743 Jena, Germany}


%\date{\today}

\begin{abstract}
Narrowband x- and gamma-ray sources based on the inverse Compton scattering of laser pulses
suffer from a limitation of the allowed laser intensity due to
the onset of nonlinear effects
that increase their bandwidth.
It has been suggested that laser pulses with a suitable frequency modulation could
compensate this ponderomotive broadening and reduce
the bandwidth of the spectral lines, which would allow to operate narrowband
Compton sources in the high-intensity regime.
In this paper we, therefore, present the theory of nonlinear Compton scattering
in a frequency modulated intense laser pulse. We systematically
derive the optimal frequency modulation of the laser pulse from the scattering matrix element
of nonlinear Compton scattering, taking into account the electron spin and recoil.
We show that, for some particular scattering angle, 
an optimized frequency modulation completely cancels the ponderomotive broadening for all harmonics
of the backscattered light. We also explore how sensitive this compensation
depends on the electron beam energy spread and emittance, as well as the laser focusing.
\end{abstract}

\maketitle

\section{Introduction}

The inverse Compton scattering of a laser pulse on a bunch of high-energy electrons
generates bright and short pulses of x-rays or gamma-rays
due to the Doppler up-shift of the laser light
\resub{\cite{Milburn:PRL1963,Arutyunian:PhysLett1963,Kulikov:PhysLett1964,%
Pantell:IEEEJQE1968,Bunkin:PhysUspekh1973,book:Fedorov,%
Schoenlein:Science1996,Kumita:LasPhys2006,Debus:APB2010,Albert:PRSTAB2010,Albert:PRSTAB2011,%
Laundy:NIMA2012,Phuoc:NatPhot2012,Du:RevScI2013,%
Corde:RevModPhys2013,Chen:PRL2013,Powers:NatPhoton2014,Sarri:PRL2014}}.
Many applications of these inverse Compton x-ray sources require a high spectral brightness,
i.e.~a large number of photons in a narrow spectral bandwidth
\cite{Carpinelli:NIMA2008,Bertozzi2008,Johnson:AIPConfProc2011,Quiter2011,Achterhold:SciRept2013}.
It is therefore essential to achieve a good control over all parameters
that contribute to the bandwidth to minimize their influence \cite{Hartemann:PRL2013,Jochmann:PRL2013,Rykovanov2014}.
In the linear regime (i.e.~the normalized laser vector potential $a_0$ is much smaller than unity)
the bandwidth is determined mainly by the bandwidth of the laser, as well as
the energy spread and emittance of the electron bunch, and the x-ray photon yield
is proportional to the laser intensity \cite{Jochmann:PRL2013}.

When the laser intensity is increased to get more x-ray photons,
an additional spectral broadening occurs:
The inhomogeneous ponderomotive force during the course of the
laser pulse changes the
electrons' longitudinal velocity and leads to a variable red-shift during the scattering such that
the scattered radiation is chirped and its spectral bandwidth increases
\cite{Hartemann1996, Brau2004, Seipt:PRA2010,Terzic:PRL2014}.
This ponderomotive broadening is a nonlinear finite-pulse-length effect
that is relevant whenever the laser bandwidth is small enough to resolve
the ponderomotive red-shift of the scattered radiation \cite{Hartemann:PRL2010,Hartemann:PRL2013}.
Thus, one needs to keep the laser intensity low enough for the bandwidth requirements,
which limits the x-ray's spectral brightness \cite{Hartemann:PRL2013,Rykovanov2014}.
Recently, it was proposed to compensate the ponderomotive broadening by using suitably chirped initial laser pulses \cite{Ghebregziabher:PRSTAB2013,Terzic:PRL2014}, which would allow to operate
the Compton sources in the high-intensity regime.
These proposals for compensating the ponderomotive broadening
were just based on the classical theory of Thomson scattering that neglects the electron recoil \cite{Ghebregziabher:PRSTAB2013,Terzic:PRL2014}.
However, there are well-known differences between the spectra of Thomson scattering and Compton scattering
due to electron recoil and spin effects \cite{Seipt:PRA2011,Boca:EPJD2011,Krajewska:LasPB2013,Krajewska:PRA2014c}.
It is, therefore, important to study whether a full compensation of the
ponderomotive broadening is still possible when these effects are taken into account.
Moreover, it was conjectured in \cite{Terzic:PRL2014}, but not proven, that an optimally
chirped frequency should remove the ponderomotive broadening from all harmonics simultaneously.

In this paper we derive the optimally chirped frequency of the initial laser pulse from the scattering matrix element of nonlinear Compton scattering within the framework of strong-field quantum electrodynamics,
fully taking into account the effect of electron recoil and spin.
We show that this optimal frequency modulation removes the ponderomotive broadening completely for a just
single scattering angle for all harmonics. For all other angles, the ponderomotive broadening is not completely compensated.
Moreover, the angular behavior of the emitted x-rays' frequency
is modified in a compensated nonlinear Compton source
and this can help to further reduce the bandwidth of the scattered radiation and to increase the photon yield.
We study the influence of the
electron beam energy spread and emittance as well as the laser beam focusing on the x-ray bandwidth
for a scenario that is realistic for electrons from a current laser plasma accelerator.

Our paper is organized as follows:
{In Section \ref{sect:theory} we present the
theory of nonlinear Compton scattering in a chirped laser pulse.
{In Section \ref{sect:compensation} we discuss the ponderomotive broadening of the spectral
lines and we derive the optimized chirping prescription for its compensation.}
We explore
the properties of the optimal frequency modulation and the compensated spectra
in Section \ref{sect:discussion}, where
we also estimate the influence of the electron beam energy spread and emittance on
the compensation.
In Section \ref{sect:summary} we summarize and conclude our results.
Some details regarding the choice of initial conditions
for the optimal frequency modulation are discussed in the Appendix.
}

%\vfill

\section{Theoretical Background}

\label{sect:theory}

Nonlinear Compton scattering (CS) is a strong field QED scattering process
where an initial electron with asymptotic four-momentum $p$
collides with a high-intensity laser pulse,
absorbs $\ell$ photons with four-momentum $k$ from the laser field, and emits
a single high-energy photon with momentum $k'$,
while the electron momentum changes during the scattering to $p'$.
The theory of strong field QED in high-intensity laser fields,
including the nonlinear Compton scattering process \cite{Nikishov:JETP1964a,Nikishov:JETP1964b,
Goldman:PhysLett1964,Brown:PR1964,Nikishov:JETP1965,Narozhnyi:JETP1965}
is based on the Furry picture, where the interaction of the electrons with the laser field $A$ is treated non-perturbatively by using laser dressed Volkov states \cite{Volkov:1935}.
These Volkov states $\Psi_{p}(x)$ are solutions of the Dirac equation %in the background field $A$,
\begin{align}
  \left( i\slashed\partial - e \slashed A + m \right) \Psi_{p}(x) = 0 \,,
  \label{eq:dirac}
\end{align}
where $e$ and $m$ are the electron charge and mass, respectively.
{We employ the Feynman slash notation $\slashed A=\gamma\cdot A = \gamma^\mu A_\mu$ for scalar products of four-vectors with the Dirac matrices $\gamma^\mu$.
Moreover, throughout the paper we use Heaviside-Lorentz units with $\hbar=c=1$,
and the fine structure constant is $\alpha=e^2/4\pi$.}

When the electrons are propagating in an infinitely extended plane-wave laser field,
their average momentum
is given by the quasi-momentum $q^\mu=p^\mu+U^\mu$ that differs from the asymptotic momentum
$p^\mu$ by the ponderomotive four-potential
\cite{Reiss:PRA2014}
\begin{align}
U^\mu =  \frac{e^2 \langle - A_\nu A^\nu \rangle }{2k\cdot p} k^\mu \,,
\label{eq:ponderomotive.potential}
\end{align}
and whose square can be interpreted as an effective mass shell condition
\cite{Ritus:JSLR1985,Reiss:PRA2014}.
The occurrence of the quasi-momentum also modifies the energy-momentum conservation
conditions for nonlinear CS, $q + \ell k = k'+q'$,
and the frequency of the scattered photons is red-shifted as compared to linear CS \cite{Ritus:JSLR1985}.
In other words, fast electrons, that face a counter-propagating laser field, do become slower due to the ponderomotive four-potential and, hence,
the Doppler up-shift is smaller.

In a pulsed laser field with a finite temporal duration
the electron's quasi-momentum and effective mass shell become
time-dependent which makes their interpretation much more sophisticated \cite{Harvey:PRL2012}.
The time-dependent quasi-momentum, for instance, causes a time-dependent red-shift of the
scattered radiation. As a consequence the scattered radiation
is chirped and it's spectral lines become much broader,
with a number of subsidiary peaks \cite{Seipt:PRA2010,Seipt:LasPhys2013}.
This phenomenon is denoted as ponderomotive broadening \cite{Krafft:PRL2004,Narozhnyi:JETP1996,Boca:PRA2009,Seipt:PRA2011}.

Other aspects of nonlinear Compton scattering in \text{short laser pulses} 
that affect inverse Compton x-ray sources include
the prospects for the generation of x-ray frequency combs \cite{Krajewska:PRA2014a,Krajewska:PRA2014b},
asymmetries in the angular distributions \cite{Mackenroth:PRL2010,Seipt:PRA2013},
spin- \cite{Boca:NIMB2012,Krajewska:LasPB2013,King:PRA2013,Krajewska:PRA2014c}
and higher-order QED effects \cite{DiPiazza:PRL2010,Seipt:PRD2012,Mackenroth:PRL2013,Ilderton:PRD2013b,Seipt:PRA2014}.
Moreover, using vortex beams that carry orbital angular momentum 
in a Compton backscattering set-up allows to
generate high-energy vortex beams \cite{Jentschura:PRL2011} or
to control the spatial distribution of the generated x-rays \cite{Seipt:PRA2014b}.
\resub{The concept of coherent laser-like x-ray sources based on stimulated inverse Compton scattering has also been investigated
\cite{Pantell:IEEEJQE1968,Bunkin:PhysUspekh1973,book:Fedorov,Harutunian:PhysRep1996,%
Avetissian:PRSTAB2007,%
Debus:APB2010,Steiniger:JPB2014}.}
For a recent review on nonlinear Compton scattering and other strong field QED processes see \cite{DiPiazza:RevModPhys2012}. In the following we extend these previous theoretical studies
by considering the nonlinear Compton scattering of a \textit{chirped} intense laser pulse.

\subsection{Vector Potential of a Chirped Laser Pulse}
\label{subsect:laserfield}

Let us consider a plane-wave laser field that propagates along
the negative z-axis.
Such a field can be described by the vector potential
\begin{align}
 A^\mu(x^+)
       &= A_0 \, g( x^+ ) \, {\rm Re} \, (\varepsilon_+^\mu e^{-i\Phi(x^+)} ) \,,
       \label{eq:vector.potential}
\end{align}
which depends only on the light-front variable $x^+ = n\cdot x = t+z$, and with the
light-like laser propagation four-vector $n^\mu=(1,0,0,-1)$.
Because we consider here chirped laser pulses with a non-constant local frequency $\omega(x^+)$,
the carrier wave depends on a general phase $\Phi(x^+)$, such that
$\omega(x^+) = \partial \Phi / \partial x^+$.
For the standard case of an unchirped wave with a
constant frequency $\omega_0$ the phase of the carrier wave is just given by $\Phi = \omega_0 x^+$.

The normalized vector potential $a_0 = {e A_0}/{m}$ is defined with respect to the peak value $A_0$.
The complex polarization vectors are defined by
$\varepsilon_\pm^\mu = \cos \eta \, \delta_1^\mu \pm i \sin \eta \, \delta_2^\mu $, with the Kronecker symbol $\delta^\mu_\nu$, and they are normalized to $\varepsilon_+\cdot \varepsilon_- = -1$.
The parameter $\eta$ describes the polarization of the laser light, where in particular $\eta=0$ ($\pi/2$) corresponds to linear (circular) polarization.

The function $g(x^+)$ describes the temporal envelope of the laser pulse. In this paper we assume a Gaussian
pulse
$g(x^+) = \exp\{ - (x^+)^2/2\Delta^2 \}$
with pulse duration $\Delta$. We moreover
assume that the pulse envelope changes slowly compared to the oscillations of the carrier wave.
We thus encounter two timescales: A fast timescale related to the oscillation of the
carrier wave $\propto 1/\omega$,
and a slow timescale related to the variation of the pulse envelope with the pulse duration $\Delta$,
such that $\omega \Delta \gg 1$.
This is a reasonable assumption in view of narrowband x-ray sources, since the laser bandwidth contribution to the x-ray spectral bandwidth
scales as $1/(\omega\Delta)$.
We moreover assume that the local frequency $\omega(x^+)$ changes also on the slow timescale.
The validity of this assumption will be verified a posteriori when we have
calculated the optimal frequency modulation for the bandwidth reduction.

\subsection{Calculation of the Matrix Element and the Differential Emission Probability}

From the Feynman rules of strong field QED in the Furry picture we find the 
$S$ matrix element for nonlinear Compton scattering
as \cite{Mitter:1975,Harvey:PRA2009}
\begin{align}
S &= \langle \mathbf p' r';\mathbf k'  \lambda' | \hat S[A] | \mathbf p r \rangle  \nonumber \\
	&=
	  -i e \intop \! \d^{4}x \: \bar{\Psi}_{p',r'}(x) \,  
	  		\slashed\varepsilon'^*_{(\lambda')} e^{i k'\cdot x} \, \Psi_{p,r}(x) \,,
\label{eq:Smatrix}
\end{align}
where $\varepsilon'^*_{(\lambda')}$ is the polarization vector of the emitted photon in the polarization state $\lambda'$.
The interaction of the electrons with the laser pulse is described non-perturbatively by using Volkov
electron wavefunctions \cite{Volkov:1935}: The positive energy solutions of the Dirac equation \eqref{eq:dirac}
can be written for a chirped plane-wave vector potential \eqref{eq:vector.potential} as
\begin{align}
 \Psi_{p,r}(x) &= \left[ 1 + \frac{e \, \slashed n \slashed A(x^+) }{2n\cdot p} \right]
  \exp\{ iS_p(x) \} u_{pr} \,,
  \label{eq:def_Ep}
\end{align}
where
\begin{align}
S_p(x) &=-p\cdot x - \frac{1}{2n\cdot p}\int \limits^{x^+}  \! \mathrm d\xi \, \big[ 2 e p\cdot A(\xi) - e^2 A^2(\xi) \big] 
\label{eq:hamilton}
\end{align}
denotes the classical Hamilton-Jacobi action
and $u_{pr}$ is the Dirac bi-spinor for a free electron with momentum $p$ in the spin state $r$,
normalized to $\bar u_{pr} u_{pr} = 2m$.
Note that the chirped frequency $\omega(x^+)$ does not appear
explicitly in the expressions for the Volkov wavefunction, \eqref{eq:def_Ep} and
\eqref{eq:hamilton}. It only enters the Volkov state via the phase factor $\Phi(x^+)$ of the vector potential \eqref{eq:vector.potential}.

The space-time integrations in \eqref{eq:Smatrix} are best performed 
using light-front coordinates, $x^\pm=x^0 \pm x^3$ and
$\mathbf x^\perp = (x^1,x^2)$, that are adapted to the light-like
laser four-direction $n^\mu=(1,0,0,-1)$, i.e.~$n^\mu$ has only one non-vanishing light-front component
$n^- =2$, while $n^+=0$ and $\mathbf n_\perp=0$.
Scalar products between four-vectors read in light-front coordinates
$x\cdot y = \frac{1}{2} x^+y^- +\frac12 {x^-}{y^+} - \mathbf x^\perp \cdot \mathbf y^\perp$.
Moreover, the four dimensional integration measure is $\d^4x =  \frac12 \d x^+ \, \d x^- \d^2 \mathbf x^\perp $.

Using these coordinates the integrations over $\d x^-$ and $ \d^2 \mathbf x^\perp$
in \eqref{eq:Smatrix} yield
delta functions ensuring the conservation of the momentum components
\begin{align}
\mathbf{p}_\perp &= \mathbf{k}'_\perp+\mathbf{p}'_\perp\,,\label{eq:momentum.conservation.perp} \\
p^+ &=k'^+ + p'^+ \,,\label{eq:momentum.conservation.plus} 
\end{align}
and the $S$ matrix element can be written as
\begin{align}
S & =- i e (2\pi)^{3}
\delta^{2}(\mathbf{k}'_\perp+\mathbf{p}'_\perp -\mathbf{p}_\perp) 
\delta(k'^+ + p'^+  - p^+ ) 
\mathrsfs M \,,
\label{eq.S_matrix.final}
\end{align}
with the scattering amplitude
\begin{multline}
\mathrsfs M(k';r,r',\lambda')
= 
   \mathrsfs T_{0}(r,r',\lambda')\mathrsfs C_{0}(k') \\
+    \mathrsfs T_{+}(r,r',\lambda')\mathrsfs C_{+}(k') 
+    \mathrsfs T_{-}(r,r',\lambda')\mathrsfs C_{-}(k') \\
+    \mathrsfs T_{2}(r,r',\lambda')\mathrsfs C_{2}(k') 
\,,
\label{eq:matrix.element} 
\end{multline}
where the electron current factors $\mathrsfs T_j$
depend on the polarization of initial and final electrons and the scattered photons:
\begin{align}
\mathrsfs T_0(r,r',\lambda') &= \bar u_{p'r'} \, \slashed \varepsilon'^*_{(\lambda')} \, u_{pr} \,, \\
 {\mathrsfs T}_{\pm}(r,r',\lambda') & =
  				ma_0 \: \bar u_{p'r'} \left( \frac{\slashed\varepsilon_{\pm}\slashed n\slashed\varepsilon'^*_{(\lambda')}}{4 n\cdot p'}
			+ \frac{ \slashed\varepsilon'^*_{(\lambda')} \slashed n\slashed\varepsilon_{\pm}}{4 n \cdot p}
			\right) u_{pr} \,, \\
{\mathrsfs T}_2(r,r',\lambda') &= 
\frac{m^2 a_0^2}{4}\frac{ n\cdot\varepsilon'^*_{(\lambda')}}{ n\cdot p' \, n\cdot p} \:
\bar u_{p'r'} \, \slashed n \, u_{pr} \,.
\label{eq.T.currents.last}
\end{align}
The integrals over the light-front time $x^+$
%\begin{widetext}
\begin{multline}
\left\{
\begin{matrix}
\mathrsfs C_0(k')\\
\mathrsfs C_\pm(k')\\
\mathrsfs C_2(k')
\end{matrix}
\right\}
= \intop\limits_{-\infty}^{\infty} \! \d x^+ \:
\exp 
	\left( i\intop^{x^+} \!\d \xi  \:\frac{k'\cdot \pi(\xi)}{n\cdot p'} \right)
  \\
 \times
 \left\{
\begin{matrix}
1\\
g(x^+) e^{\mp i \Phi(x^+)} \\
g^2(x^+) ( 1+ {\cos2\eta \, \cos 2\Phi(x^+)})
\end{matrix}
\right\}
	\,
\label{eq:def_C}
\end{multline}
determine the dynamics of the scattering process.
%\end{widetext}
%
The function $\mathrsfs C_0$ is an infinite integral over a pure phase and is evaluated by using the
Boca-Florescu transformation \cite{Boca:PRA2009}.
The term in the exponent of the dynamic integrals \eqref{eq:def_C} contains
the classical kinetic four-momentum of the electron \cite{Nikishov:JETP1964a}
\begin{align}
\pi^\mu(x^+) = p^\mu -e A^\mu + n^\mu \frac{eA\cdot p}{n\cdot p} - n^\mu \frac{e^2A\cdot A}{2n\cdot p} \,,
\end{align}
as solution of the classical Lorentz force equation
\begin{align}
\frac{\d \pi^\mu}{\d \tau} = \frac{e}{m} F^{\mu \nu} \pi_\nu \,,
\label{eq:lorentz_force}
\end{align}
where $\tau$ is the electron's proper time and $F_{\mu\nu} = \partial_\mu A_\nu - \partial_\nu A_\mu$ is
the electromagnetic field strength tensor.
The classical equation of motion \eqref{eq:lorentz_force} can be easily integrated
since for a plane-wave laser field as in \eqref{eq:vector.potential}
the proper time $\tau$ is related to the light-front time via
$ x^+ = \frac{n\cdot p}{m} \, \tau$ \cite{Meyer:PRD1971,Heinzl:OptCommun2009}.
Moreover, the kinetic four-momentum
is the expectation value of the kinetic momentum operator
$\Pi^\mu \equiv i\partial^\mu - eA^\mu$
in a Volkov state
$ \bar \Psi_{p,r}(x) \Pi^\mu \Psi_{p,r}(x) = \pi^\mu(x^+) \,  \bar \Psi_{p,r}(x) \Psi_{p,r}(x) $
\cite{Ritus:JSLR1985}.

We can use the scattering amplitude \eqref{eq:matrix.element} to express
the energy and angular differential emission probability for a photon with
four-momentum $k'$ in nonlinear Compton scattering of chirped
laser light on unpolarized electrons as \cite{Seipt:PRA2011}
\begin{align}
\frac{\d N}{\d \Omega \d \omega'} &= \frac{1}{2} \sum_{\lambda',r,r'}
 \frac{e^2 \omega'}{64 \pi^3 \: n\cdot p' \: n\cdot p} 
|\mathrsfs M(k';r,r',\lambda')|^2 \,,
\label{eq:probability}
\end{align}
if we assume that the polarization of the scattered electron and photon remains unobserved.
The four-momentum of the scattered photon is determined by the frequency $\omega'$ and the direction $\mathbf n'(\vartheta,\varphi) = ( \cos \varphi \, \sin \vartheta , \sin \varphi \, \sin \vartheta ,  \cos \vartheta)$
via $k' =  \omega' n' = \omega' (1,\mathbf n')$.
Note that all components of the final electron momentum $p'$ are fixed by
the light-front momentum conservation conditions \eqref{eq:momentum.conservation.perp} and \eqref{eq:momentum.conservation.plus}, and by the asymptotic light-front mass-shell condition
$p'^- = (\mathbf p_\perp^2 + m^2)/p'^+$.

\subsection{Separation of Slow and Fast Time Scales and Higher Harmonics}

The energy and angular differential emission probability, Eq.~\eqref{eq:probability}, describes the
spectrum of emitted x-rays for an arbitrarily chirped laser pulse,
including the emission
of higher harmonics and the effect of ponderomotive broadening due to the gradual slow-down
of the electron as it enters the high-intensity regions at the peak of the laser pulse
\cite{Boca:PRA2009,Seipt:PRA2011}.
To find the optimal frequency modulation of the
initial laser pulse $\omega(x^+)$ that compensates the ponderomotive broadening we need an expression
for the scattering matrix element where we explicitly see how the slow-down of the electron, which
happens on the time-scale of the pulse envelope, affects the shape of the spectral lines of the emitted x-rays.
We thus need to separate the slow (pulse duration $\Delta$) and fast (carrier wave period $1/\omega$)
time-scales of the electron kinetic four-momentum $\pi^\mu(x^+)$ that appears in the exponent of the dynamic integrals \eqref{eq:def_C}. This separation of time-scales is only meaningful for laser pulses that contain many oscillations of the carrier wave, i.e.~for $\omega \Delta \gg 1$.

To achieve the separation of time-scales we employ a suitable floating average with a variable local window size,
\begin{align}
\langle f \rangle(x^+)
= \frac{\omega(x^+)}{2\pi} \int_{x^+-\pi/\omega(x^+)}^{x^++\pi/\omega(x^+)} \d \xi \, f(\xi)\,,
\end{align}
that averages over the fast oscillations of the carrier wave,
such that, e.g.,~$\langle A^\mu \rangle = 0$ and $\langle g^2 \rangle(x^+) = g^2(x^+)$ within the slowly varying envelope approximation, i.e.~assuming that $g$ and $\omega$ are approximately constant over
the averaging window of one period of the carrier wave \cite{Narozhnyi:JETP1996}.
The averaged electron kinetic four-momentum (i.e.~the quasi-momentum)
$q^\mu(x^+) = \langle \pi(x^+) \rangle = p^\mu + U^\mu(x^+)$
depends explicitly on (light-front) time, but only on the slow time-scale of the pulse envelope via the ponderomotive four-potential
\begin{align}
U^\mu(x^+) =  \frac{m^2 a_0^2 }{4 n\cdot p} \, g^2(x^+) \,n^\mu  \,.
\end{align}
Moreover, the value of the ponderomotive four-potential is \textit{independent} of the laser polarization
due to our choice of normalized polarization vectors $\varepsilon_+\cdot \varepsilon_- = -1$ which allows a unified treatment of arbitrary laser polarization \footnote{Note that often the ponderomotive potential is defined unequally for linear and circular laser polarization.}.

We can use the quasi-momentum to define
the oscillating part of the classical kinetic four-momentum, $ \pi^\mu_{\rm osc}(x^+) := \pi^\mu(x^+) - q^\mu(x^+)$, that oscillates with the frequency $\omega$ of the carrier wave,
and that averages to zero, $\langle \pi^\mu_{\rm osc}(x^+) \rangle =0$, within the slowly varying envelope approximation \cite{Narozhnyi:JETP1996}.

The fast oscillating term $\pi^\mu_{\rm osc}$ leads, as we see below, to the formation of higher
harmonic lines in the frequency spectrum of the generated x-rays. In contrast, the quasi-momentum $q^\mu(x^+)$ determines the shape of each of these lines by controlling the ponderomotive broadening due to the slow longitudinal dephasing of the electron motion in the laser field.

We now disentangle these two effects by using the above separation of time-scales
in the expression \eqref{eq:def_C} for the dynamic integrals $\mathrsfs C_j$,
that enter the scattering amplitude \eqref{eq:matrix.element}.
This allows, in the end, to determine the optimal frequency modulation $\omega(x^+)$ of the initial
laser pulse to compensate the ponderomotive broadening of the spectral lines.
For convenience, let us specify a circularly polarized laser (with $\eta=\pi/2$)
in the following.
Defining the complex coefficient
 \begin{align}
 \alpha_+ &= \frac{ma_0}{n\cdot p'} \left( {k'\cdot \varepsilon_+}
 - \frac{k'\cdot n \: p\cdot \varepsilon_+}{n\cdot p}
 \right) \,,
 \label{eq.def.alpha}
\end{align}
that can be represented by its magnitude $\bar \alpha=|\alpha_+|$ and phase $\phi_\alpha$ as $\alpha_+ = \bar \alpha \, e^{i\phi_\alpha}$,
we can write for the oscillating part
of the electron kinetic four-momentum in the exponent of \eqref{eq:def_C}
\begin{align}
i  \int \! \d \xi \: \frac{k'\cdot \pi_{\rm osc}( \xi )}{n\cdot p'} 
= - i  \, {\rm Re}  \left(  \alpha_+ \: \int \! \d \xi \: g( \xi) e^{-i\Phi( \xi)} \right) \,.
\label{eq:int.osc}
\end{align}
Moreover, the integral on the right hand side of Eq.~\eqref{eq:int.osc}
is evaluated by an integration by parts
\begin{align}
\int \! \d \xi \: g(\xi) e^{-i\Phi(\xi)}  
	\approx  i \, \frac{g(\xi)}{\omega(\xi)} \, e^{ - i \Phi(\xi) } %+ \mathcal O(1/\Delta)
	\label{eq:SVEA}
\end{align}
with the local frequency $\omega(\xi) = \frac{\partial \Phi(\xi)}{\partial \xi}$, where
we only keep the surface term within the slowly varying envelope approximation \cite{Seipt:PRA2011}.

These results allow us to apply the Jacobi-Anger expansion \cite{book:Watson}
to the oscillating part of the laser kinetic momentum in the exponents of
\eqref{eq:def_C}, yielding
\begin{multline}
e^{i \int^{x^+} \!\d \xi  \: \frac{k'\cdot \pi_{\rm osc}(x^+)}{n\cdot p'} } 
= e^{ - i \bar \alpha \frac{g(x^+)}{\omega(x^+)} \sin \left( \Phi(x^+) - \phi_\alpha \right)} \\
= \sum_{\ell=-\infty}^\infty J_\ell \left( \bar \alpha \frac{g(x^+)}{\omega(x^+)} \right) 
e^{-  i \ell  \Phi(x^+)}  e^{ i  \ell \phi_\alpha  } \,,
\label{eq:harmonics}
\end{multline} 
where the coefficients $J_\ell$ are Bessel functions of the first kind \cite{book:Watson}.
This expression is the expansion into a series of harmonics that are interpreted as the net absorption of a number of $\ell$ laser photons from the laser field under the emission of a single photon with frequency $\omega'$.
Plugging \eqref{eq:harmonics} into \eqref{eq:def_C}
provides the harmonic expansion of the dynamic integrals that enter the scattering amplitude as
\begin{widetext}
\begin{align}
\left\{
\begin{matrix}
\mathrsfs C_0 \\
\mathrsfs C_\pm \\
\mathrsfs C_2
\end{matrix}
\right\}
= \sum_{\ell=1}^\infty e^{i\ell \phi_\alpha} \intop_{-\infty}^\infty  \!\d x^+ 
\left\{
\begin{matrix}
J_\ell \left( \frac{\bar \alpha g(x^+)}{\omega(x^+)} \right) \\
g(x^+)  J_{\ell\pm1} \left( \frac{\bar \alpha g(x^+)}{\omega(x^+)} \right) e^{\pm i \phi_\alpha} \\
g^2(x^+) J_\ell \left( \frac{\bar \alpha g(x^+)}{\omega(x^+)} \right) 
\end{matrix}
\right\}
 e^{ i \intop^{x^+} \! \d \xi\: 
		\left( \frac{k'\cdot q(\xi )}{n\cdot p'} - \ell \omega(\xi)\right)
		}
		\,,
		\label{eq:A_expanded}
\end{align}
\end{widetext}
where the averaged kinetic momentum $q^\mu$ in the exponent just represents
the ``slow part'' of the electron motion.
Moreover, the expression \eqref{eq:A_expanded} is suitable to derive the conditions for a compensation
of the ponderomotive broadening of the harmonic lines, i.e.~the precise
form of the laser frequency modulation $\omega(x^+)$ that remained undefined up to now.

In the expansions \eqref{eq:harmonics} and \eqref{eq:A_expanded}, the coefficients $J_\ell$ determine
the strength of the emission into the $\ell$-th harmonic line of the emitted x-ray spectrum.
Their dependence on $x^+$ indicates that different harmonics are generated at different times during the course of the pulse, e.g.~very high harmonics are generated only at the center of the pulse where the laser intensity---and therefore the argument of the Bessel functions---is largest.
For linear laser polarization, we could obtain a similar expansion as \eqref{eq:harmonics},
but with the expansion coefficients as two-argument Bessel functions
\cite{Loetstedt:PRE2009,Leubner:PRA2011}.
While the pre-exponential in the brackets of \eqref{eq:A_expanded} differ,
we note again that its exponential part
with the quasi-momentum $q^\mu$ is equal for linear and circular laser polarization.

\section{Ponderomotive Broadening and its Compensation by Laser Frequency Modulation}

\label{sect:compensation}

In this section we explicitly show how the ponderomotive broadening of the spectral lines of the emitted x-rays
arises, and we derive the optimal frequency modulation of the initial laser pulse that removes the
ponderomotive broadening of the spectral lines completely.

The positions of the spectral lines of the emitted radiation are determined by those times at which the exponent of the dynamic integrals in \eqref{eq:A_expanded} becomes stationary.
Note that each of the terms in the expansion \eqref{eq:A_expanded} possesses two
real stationary points, in contrast to the non-expanded integrals \eqref{eq:def_C} for which the stationary point lie off the real axis \cite{Mackenroth:PRA2011,Seipt:LasPhys2013}.
The stationarity condition
\begin{align}
\frac{k'\cdot q(x^+)}{n\cdot p -n\cdot k'} - \ell \omega(x^+) = 0
\end{align}
determines the \textit{local} value of the scattered photon's frequency for each harmonic $\ell$ as
\begin{align}
\omega'(x^+) = \frac{\ell \omega(x^+) n\cdot p}{n'\cdot p + n'\cdot U(x^+) + \ell \omega(x^+) n'\cdot n} \,,
\label{eq:omega.SPA}
\end{align}
where $n'$ denotes the four-direction of the emitted photon $k' = \omega' n'$.
{Here we derived the frequency of a photon that is emitted at a specific time $x^+$ during the course of
the pulse. Its time dependence stems from the time-dependent ponderomotive potential $U^\mu(x^+)$
and, since we are treating chirped laser pulses, also from the frequency modulation $\omega(x^+)$.
}

Let us first consider the standard case of an unchirped laser pulse and see what this implies for the time-dependence of the scattered radiation's frequency.
For an unchirped laser field with the constant frequency $\omega(x^+) = \omega_0$ we see that
the scattered radiation is chirped and its frequency
\begin{align}
\omega'(x^+) 
= 
\frac{\ell \Omega}{1 +\ell \chi + \beta_0 g^2(x^+) }
	\label{eq:omega.SPA.0}
\end{align}
varies between the linear Compton lines $\omega' = \ell \Omega/(1+\ell \chi)$
and the maximally red-shifted Compton lines $\omega' = \ell \Omega/(1+\ell \chi + \beta_0)$.
Here, $\Omega=\omega_0 \frac{ n\cdot p}{n'\cdot p}$ is the frequency of the fundamental line in linear Thomson scattering, i.e.~if the electron recoil $\chi = \omega_0 \frac{ n\cdot n' }{n'\cdot p}$ is neglected.
Moreover, the value of 
\begin{align}
\beta_0=\frac{m^2 a_0^2}{4} \frac{n'\cdot n}{n'\cdot p\: n\cdot p}
\label{eq:def.beta0}
\end{align}
determines the maximum red-shift of the scattered radiation, i.e.~the effect of the ponderomotive slow-down of the electron
as it enters the high-intensity part of the laser pulse.
The integration over the pulse in \eqref{eq:A_expanded} collects
all these frequency components, and the ponderomotively broadened spectral lines
just localize between the linear and maximally red-shifted Compton lines.
The ponderomotively broadened lines consist of a number of subsidiary peaks due to
interference of two stationary phase points, with the largest peak
close to the maximally red-shifted Compton line \cite{Seipt:LasPhys2013}.

Let us now turn back to the case of a chirped laser pulse.
In particular, we now determine a specific prescription for $\omega(x^+)$ in order to compensate the ponderomotive broadening described above.
The condition to achieve this can be formulated as follows: During the course of the laser pulse the emitted photon frequency $\omega'(x^+)$ should be constant, i.e.~we require the time-derivative of Eq.~\eqref{eq:omega.SPA} to vanish:
\begin{align}
\frac{\partial \omega'(x^+)}{\partial x^+} &= 0 \,.
\label{eq:compensation.condition}
\end{align}
This condition furnishes a differential equation
for the optimal chirp of the laser frequency $\omega(x^+)$:
\begin{align}
\frac{\dot \omega }{\omega}
=
\frac{n'\cdot \dot q}{n'\cdot q}  \,,
\label{eq:compensation.dgl}
\end{align}
where the dot means the derivative with respect to the light-front time $x^+$.
Its solution is given by
\begin{align}
\omega(x^+)
=    \omega(x_i^+) \frac{1 + \frac{n'\cdot U(x^+)}{n'\cdot p}  }{1 + \frac{n'\cdot U(x_i)}{n'\cdot p}}
=    \omega(x_i^+) \frac{1 + \beta_0 g^2(x^+)  }{1 + \beta_0 g^2(x^+_i) }
\label{eq:compensation.solution}
\end{align}
with some initial time $x^+_i$.
This chirping prescription provides the optimal frequency modulation to compensate the ponderomotive broadening of the Compton backscattered x-ray spectra.
It states how the laser frequency 
should increase during the course of the pulse
in order to counterbalance the gradual intensity dependent red-shift
due to the ponderomotive slow-down of the electrons.
Equation \eqref{eq:compensation.solution} relates the local laser frequency $\omega(x^+)$ to the laser pulse envelope $g(x^+)$ such that our initial assumption in Section \ref{subsect:laserfield} that $\omega(x^+)$ changes on the same slow time scale as $g(x^+)$ is fulfilled.
It is remarkable that the optimal frequency modulation is not influenced by the electron recoil,
as Eq.~\eqref{eq:compensation.solution} does not depend on the electron recoil parameter $\chi$,
while the equation for the local x-ray frequency
\eqref{eq:omega.SPA} that is used to define the compensation condition does depend on the recoil.

The choice of the initial conditions $x_i^+$ fixes the overall frequency scale of the chirped frequency $\omega(x^+)$ for comparison with the case of an unchirped laser pulse with frequency $\omega_0$.
We propose to fix the initial conditions at asymptotic times $x^+_i\to -\infty$ where the ponderomotive
potential vanishes as $\omega(-\infty)=\omega_0$ such that
the frequency modulation reads
\begin{align}
\omega(x^+) = \omega_0 \left( 1 + \beta_0 g^2(x^+) \right) \,.
\label{eq:optimal.chirp.asymptotic}
\end{align}
With this choice, the compensated lines condense at the linear Compton lines at $\omega'=\ell \Omega/(1+\ell \chi)$.
This allows to easily compare the compensated nonlinear
Compton source with the case of a linear Compton source with unchirped laser frequency $\omega_0$.
A more detailed discussion of our initial conditions in comparison to the ones used in the literature can be found in the Appendix.

\section{Discussion}

\label{sect:discussion}

\subsection{Properties of the Optimal Frequency Modulation}

For our discussion let us specify a typical set-up for an inverse Compton backscattering x-ray source:
We assume the laser pulse to collide head-on (i.e.~with an incidence angle of $\unit{180}{\degree}$) with
an ultra-relativistic electron bunch with four-momentum $p=(m\gamma,0,0,m\sqrt{\gamma^2-1} )$,
and a Lorentz factor $\gamma \gg 1$.
Since most of the photons are scattered
within a small $1/\gamma$ cone around the initial electron beam direction \cite{book:Jackson},
the local frequency of the scattered photons \eqref{eq:omega.SPA}
can be approximated as
\begin{align}
\omega'(x^+) = \frac{4\gamma^2 \omega(x^+) \ell}{1 + \gamma^2\vartheta^2 + 4\ell \gamma \frac{\omega(x^+)}{m} + \frac{a_0^2}{2}g^2(x^+)}
\label{eq:omega.SPA.discussion}
\end{align}
for small photon scattering angles $\vartheta$.
%(that is measured from the initial electron beam direction).
%

The influence of the ponderomotive electron slow-down on the maximum frequency red-shift of the backscattered radiation, $\beta_0 = a_0^2/(2+2\gamma^2\vartheta^2)$,
decreases with increasing scattering angle.
{Therefore,
it is impossible to remove the ponderomotive broadening from
all scattering angles at the same time.}
We have to decide at what scattering angle we want to
compensate the ponderomotive broadening, where the increase of the laser frequency exactly
balances the effect of the slow-down of the electrons.
Let us call this the optimization angle $\vartheta_0$.
For the head-on collision this optimization angle dependence of the optimal frequency modulation is just given by
\begin{align}
\omega(x^+)
=
\omega_0 \left( 1 + \frac{1}{1 + \gamma^2\vartheta_0^2 }
\frac{ a_0^2}{2} g^2(x^+)\right) \,.
\label{eq:optimal.theta0}
\end{align}
In particular, when we optimize for on-axis radiation, $\vartheta_0=0$,
our chirping prescription \eqref{eq:optimal.theta0}
coincides with the result of Ref.~\cite{Terzic:PRL2014} that was found
within a classical model neglecting the electron recoil
(apart from a different global frequency scale that can be absorbed
into a different choice of initial conditions. For a discussion
see the Appendix).
This shows again the remarkable fact that the electron recoil during the scattering does not influence
the form of the optimal frequency modulation.

\begin{figure*}[!th]
\includegraphics[width=0.99\textwidth]{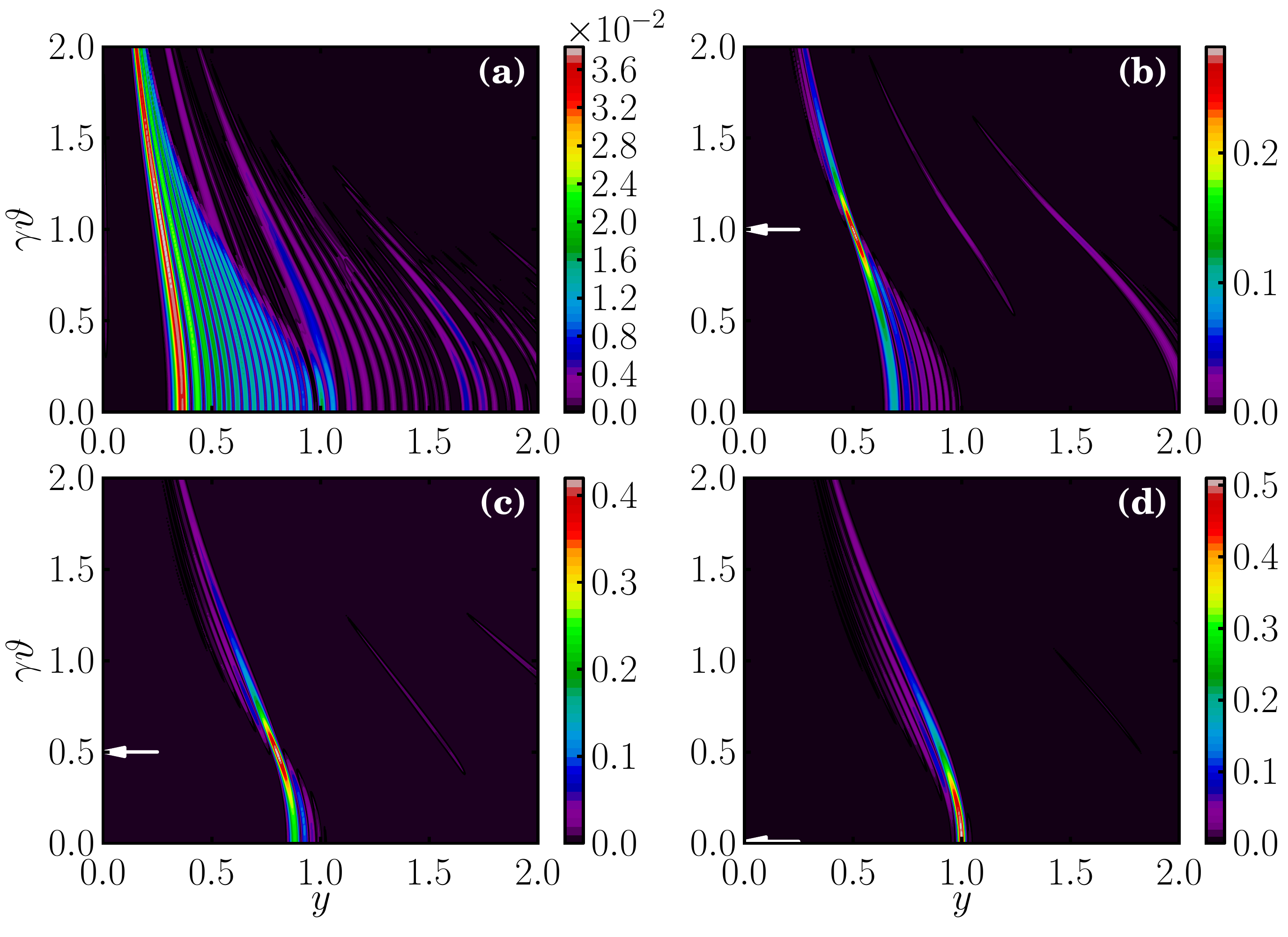}
\caption{(Color online) Compensated energy and angular differential emission probability for nonlinear Compton scattering $\d N/\d\Omega \d \omega'$, Eq.~\eqref{eq:probability}, as a function of the normalized scattering angle $\gamma\vartheta$ and normalized frequency of the scattered photons $y = {\omega'}/{4\gamma^2\omega_0}$.
We consider a high-intensity laser pulse with $a_0=2$,
frequency $\omega_0=\unit{1.55}{\electronvolt}$ and a Gaussian pulse duration of $\Delta = \unit{13}{\femto\second} $ that collides head-on with an electron beam with an energy of $\unit{51}{\mega\electronvolt}$.
The spectrum is observed
perpendicular to the linear laser polarization.
The uncompensated nonlinear CS spectrum in (a) shows the typical signature of ponderomotive
broadening with very broad spectral lines with a series of sub-peaks.
In contrast, the compensated spectra in (b)-(d) are much narrower, with the lowest spectral width
and the largest peak height occurring at the scattering angle that equals the optimization angle $\vartheta_0$, which is marked with an arrow in each panel:
$\vartheta_0=1/\gamma$ in (b), $\vartheta_0=0.5/\gamma$ in (c), and $\vartheta_0=0$ in (d).
}
\label{fig:contour}
\end{figure*}

The optimal frequency modulation removes the ponderomotive broadening from \textit{all} harmonics $\ell$
at the optimization angle $\vartheta_0$ simultaneously.
This we conclude from the fact that the chirping prescription
\eqref{eq:optimal.theta0}, does not depend on the harmonic number $\ell$, although we
derived it from the general expression for the frequency of the $\ell$-th harmonic, Eq.~\eqref{eq:omega.SPA}.

By plugging \eqref{eq:optimal.theta0} into \eqref{eq:omega.SPA.discussion} we find that
the optimally compensated harmonic lines (i.e.~for $\vartheta=\vartheta_0$) are located at the normalized frequencies
\begin{align}
y(\vartheta) = \frac{\ell}{1+\gamma^2\vartheta^2 + 4\gamma \ell \frac{\omega_0}{m} } \,.
\label{eq:y_compensated}
\end{align}
Following Refs.~\cite{Krafft:PRL2004,Terzic:PRL2014,Rykovanov2014}, the x-ray frequency $\omega'$ is conveniently normalized to the on-axis x-ray frequency in recoil-free linear Thomson scattering as $y = \omega'/4\gamma^2\omega_0$.

A numerical example for the compensation of the ponderomotive broadening and the narrowing of the spectral lines is exhibited in Fig.~\ref{fig:contour} for a laser strength of $a_0=2$.
The uncompensated spectrum in Fig.~\ref{fig:contour} (a),
where the initial laser pulse is unchirped with a frequency $\omega_0=\unit{1.55}{\electronvolt}$,
acts as a benchmark to quantify the effect of the bandwidth reduction in Fig.~\ref{fig:contour} (b)-(d).
In Fig.~\ref{fig:contour} (a) we see the strong influence of the ponderomotive broadening for $a_0=2$,
where the harmonic lines consist of many sub-peaks and are overlapping in the spectral domain,
forming a broad continuum instead of narrow lines.
The compensated nonlinear CS spectra are depicted in Fig.~\ref{fig:contour} (b)-(d), where
the chirping prescription \eqref{eq:optimal.theta0} is adjusted such that the ponderomotive broadening is
removed completely at the optimization angles $\vartheta_0= 1/\gamma$ (b), $\vartheta_0= 0.5/\gamma$ (c), and $\vartheta_0=0$ (d), marked by arrows.
The spectral peaks are much narrower than in the uncompensated case in (a), with the narrowest
bandwidth occuring at the scattering angles that coincide with the optimization angle.

For all scattering angles offside the optimization angle the ponderomotive broadening is just partially
compensated. The spectral lines are much narrower than in the uncompensated case, but can be still
considerably broader than at the optimization angle. For instance in (b) the spectrum at $\vartheta=0$
still is pretty broad with a number of spectral sub-peaks.
While the ponderomotive broadening can be removed only at one particular scattering angle $\vartheta= \vartheta_0$, the numerical results in Fig.~\ref{fig:contour} (b)-(d) show that the spectral bandwidth is considerably reduced also for scattering angles slightly off the optimization angle $\vartheta \approx \vartheta_0$.
By comparing, e.g., (a) and (b), we also see the higher harmonic lines to
become much narrower due to the compensation as predicted above.

\begin{figure}[!th]
\includegraphics[width=0.99\columnwidth]{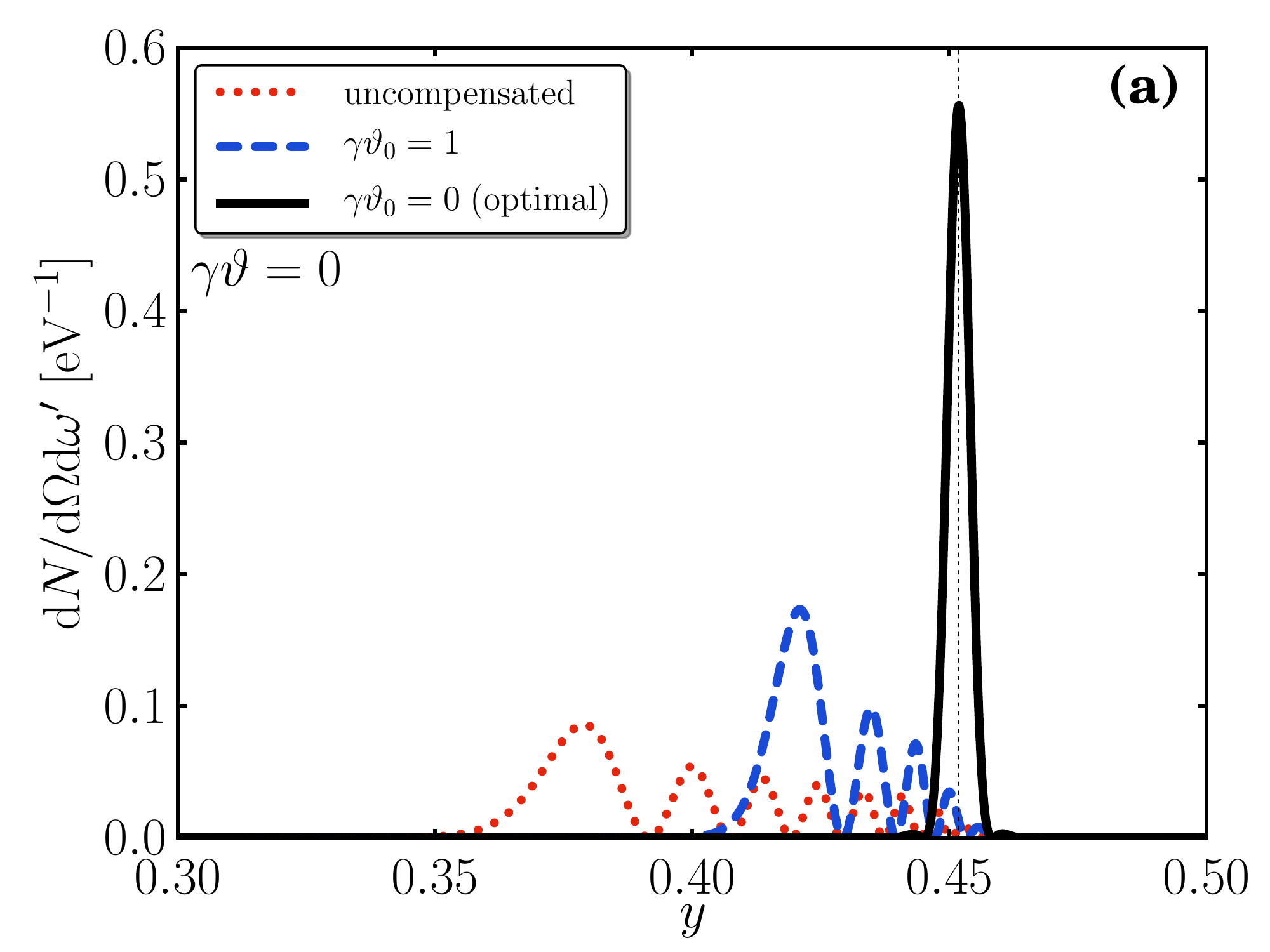}
\includegraphics[width=0.99\columnwidth]{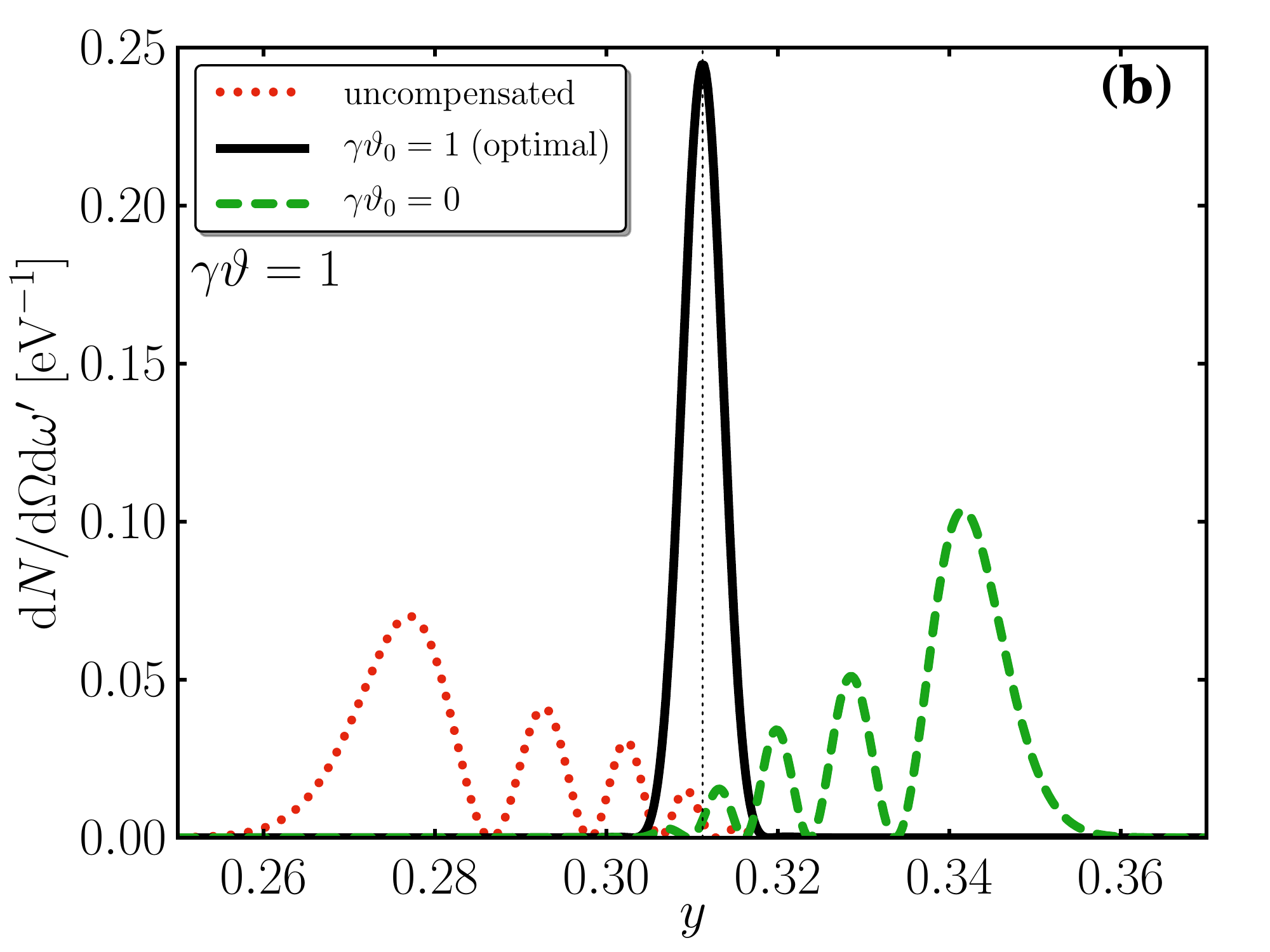}
\caption{(Color online) Detail of the frequency spectrum of the
fundamental line ($\ell=1$) for compensated nonlinear CS at fixed scattering angles
$\vartheta=0$ and $\vartheta=1/\gamma$ in (a) and (b), respectively.
The calculation is for the head-on scattering of a linearly polarized laser with pulse duration $\Delta = \unit{21}{\femto\second}$ and $a_0=1$ on an electron beam with initial energy $\unit{51}{\giga\electronvolt}$.
The optimally compensated fundamental lines (black solid curves) for $\vartheta=\vartheta_0$ are much
narrower as compared to the uncompensated spectra (red curves) and
partially compensated spectra (blue and green dashed curves, respectively), where the optimization angle $\vartheta_0$ does not coincide with the scattering angle $\vartheta$.
\vspace*{-0.1ex}
}
\label{fig:cuts}
\end{figure}

Figure~\ref{fig:cuts} demonstrates the compensation of the ponderomotive broadening
for electrons with an initial energy of $\unit{51}{\giga\electronvolt}$, i.e.~a Lorentz factor of $\gamma= 10^5$,
where the electron recoil is not negligible.
The black curves in each panel correspond to the case of an optimal compensation
of the ponderomotive broadening according to
the prescription \eqref{eq:optimal.theta0} with $\vartheta=\vartheta_0$.
Due to the effect of the electron recoil their location is shifted to lower frequencies (e.g.~$y=0.45$ instead $y=1$ for $\vartheta=0$ in (a)).
The blue dashed curve in (a) shows a non-complete compensation of the ponderomotive broadening
for $\vartheta_0>\vartheta$.

In Fig.~\ref{fig:cuts} (b), the green dashed curve shows the effect of over-compensation
of ponderomotive broadening for scattering angles larger than the optimization angle $\vartheta>\vartheta_0$.
In this case the largest peak of the spectral line appears blue-shifted compared to the optimally compensated line.
This happens because for $\vartheta_0<\vartheta$ the laser frequency increases more as is necessary to compensate the frequency red-shift due to the slow-down of the electron.

\subsection{Angular Properties of the Radiation and Increased Collimation Angle}
\label{subsect:monochromatization}

In the last section we encountered the effect of over-compensation of the spectral lines for scattering
angles $\vartheta$ that are larger than the optimization angle $\vartheta_0$.
This overcompensation also influences the angular behavior of the emitted x-rays' frequency.
Here we explore how this effect can help to increase the photon yield when optimizing for $\vartheta_0=0$,
and restricting the discussion to the fundamental line $\ell=1$.

\begin{figure}[!ht]

\includegraphics[width=0.49\columnwidth]{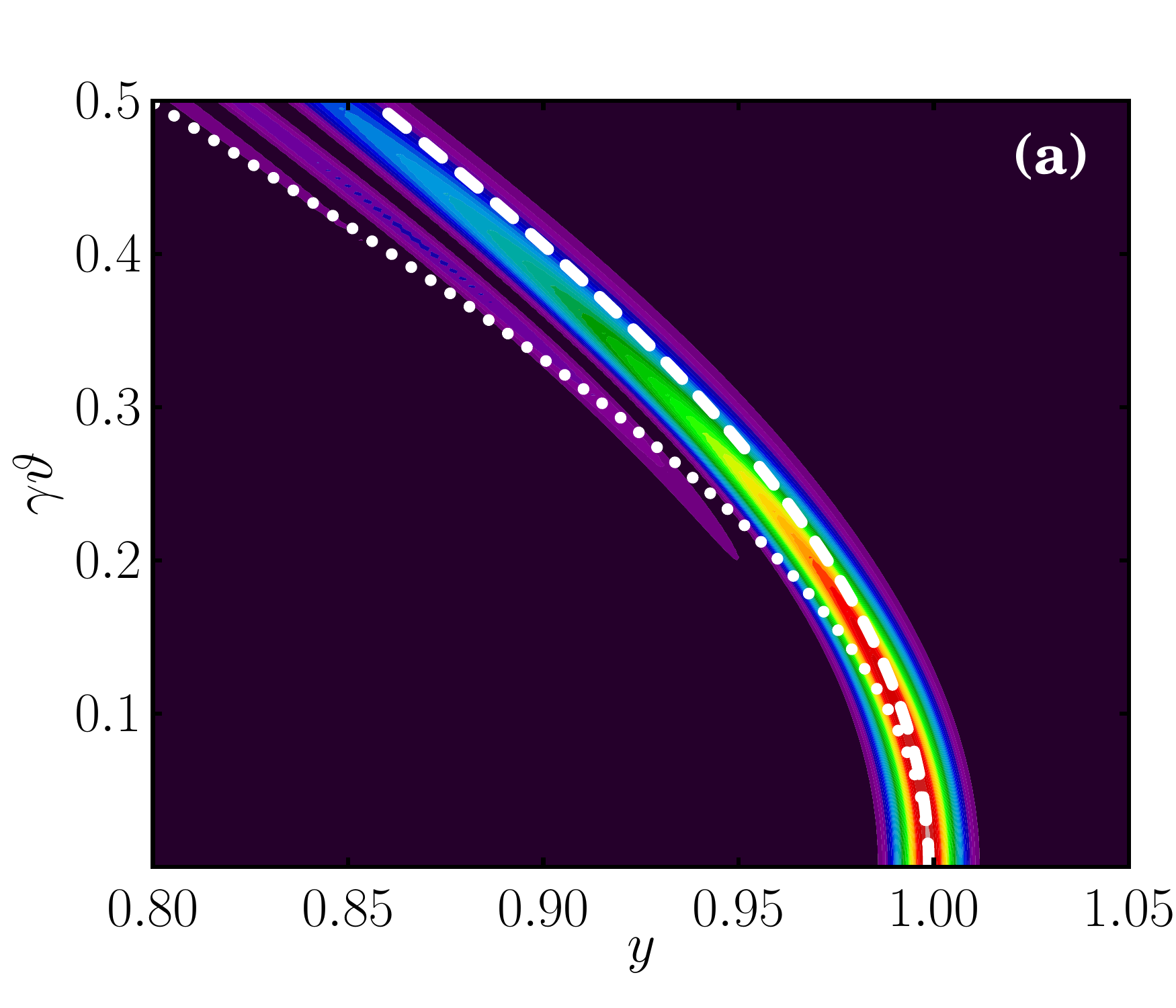}
\includegraphics[width=0.49\columnwidth]{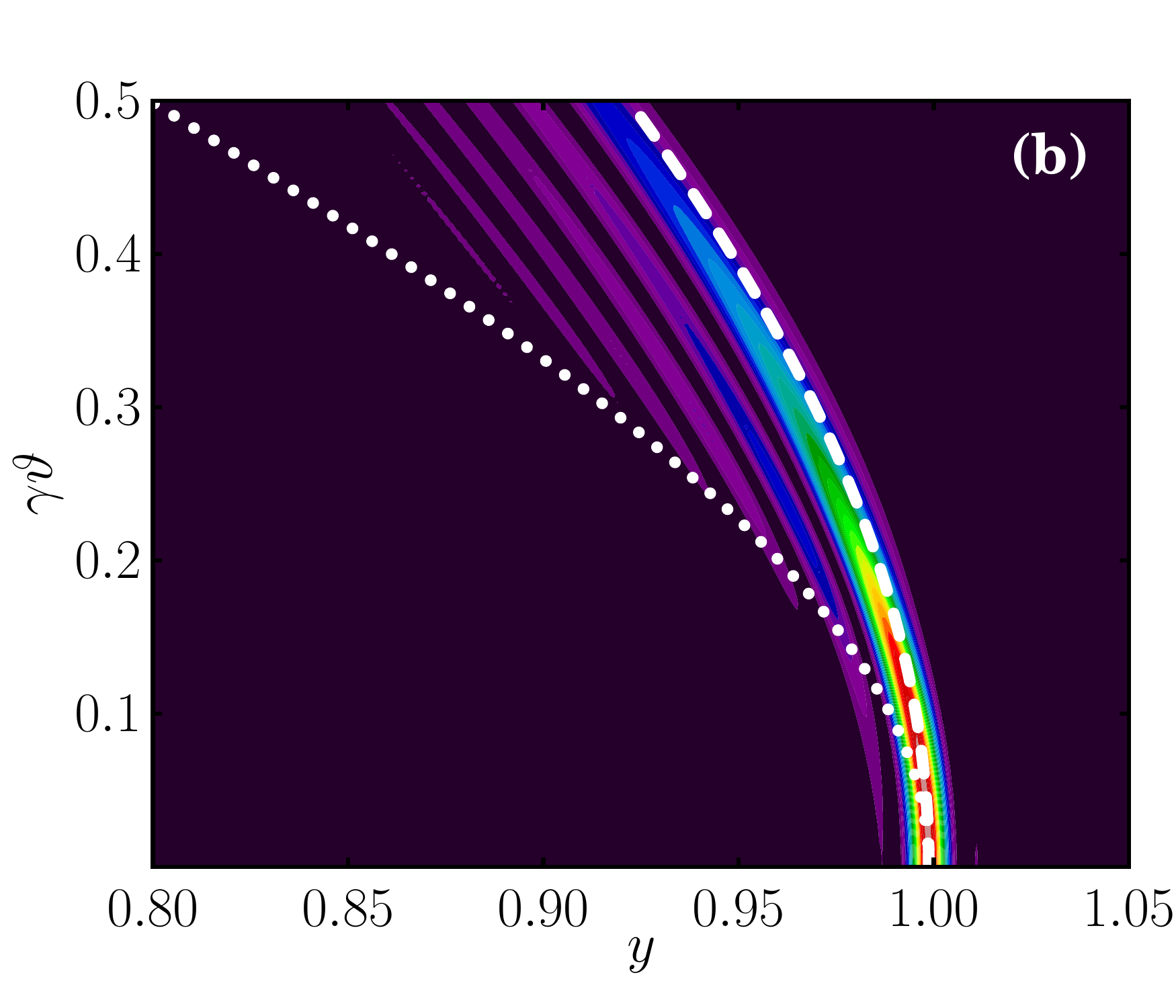}
\includegraphics[width=0.99\columnwidth]{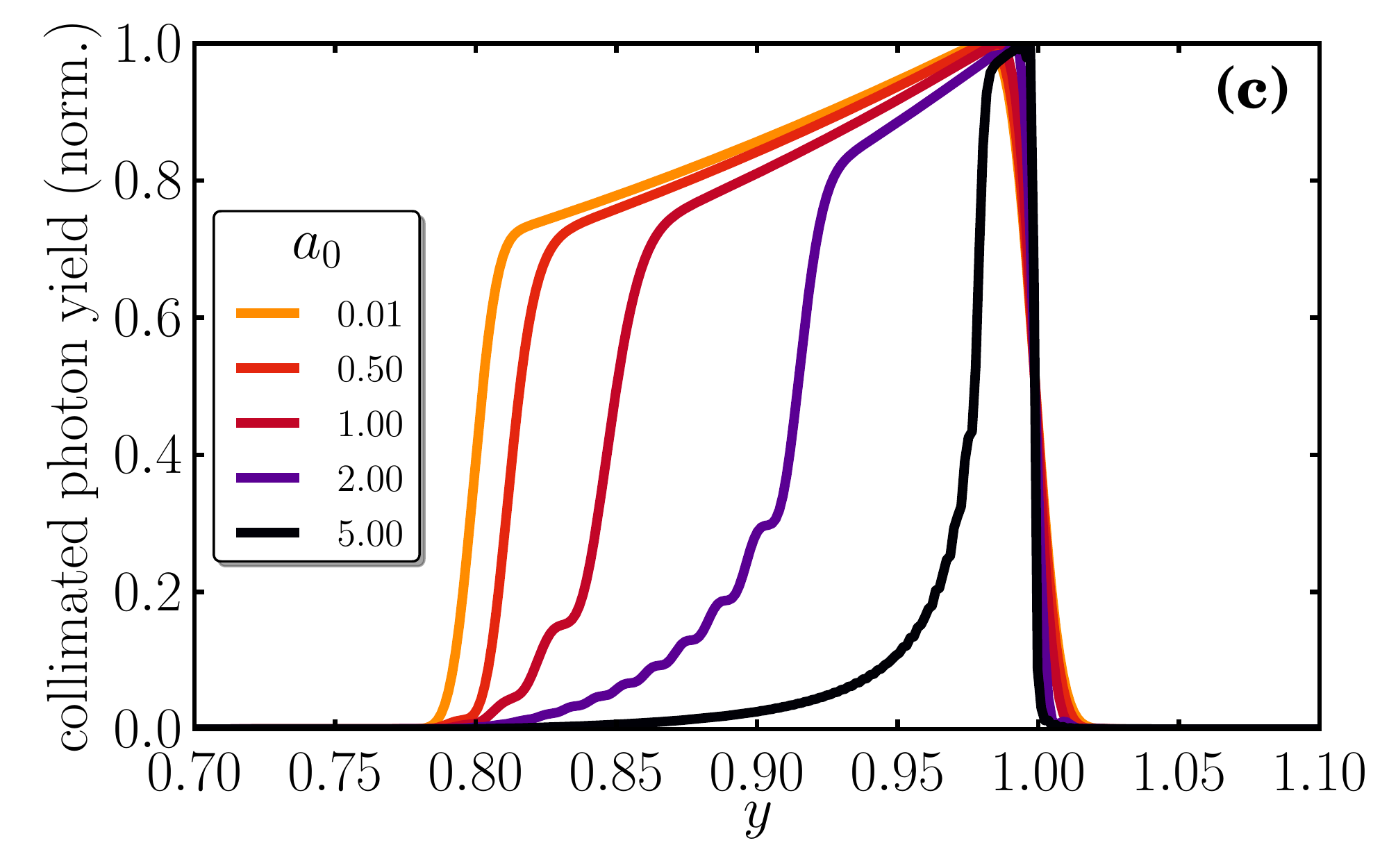}
\caption{(Color online) The normalized frequency $y$ of the largest peak of the compensated fundamental line depends much weaker on the normalized scattering angle $\gamma\vartheta$ for increasing values of $a_0$, e.g.~for $a_0=1$ (a) and $a_0=2$ (b),
and follows the modified angular dependence of the normalized photon frequency (dashed curve)
that is given by Eq.~\eqref{eq:y_gamma_star}.
The radiation becomes more monochromatic over a larger angular region, as compared to the case of
a linear CS, whose frequency-angular correlation is depicted by the dotted curve.
Thus, a high laser intensity 
$a_0 \gtrsim1$ helps to reduce the spectral bandwidth of the compensated nonlinear CS x-ray source when the scattered radiation is collected over a finite collimation angle, e.g.~$\gamma\vartheta_c=0.5$
(c).
}
\label{fig:collimation}
\end{figure}

During the interaction of an electron with the laser pulse, the electron is slowed down, i.e.~its Lorentz factor becomes effectively smaller, with the minimum given by $\gamma_\star = \gamma/\sqrt{1+a_0^2/2}$.
As the opening angle of the radiation cone that is generated by a relativistic electron is proportional to $1/\gamma$, one can expect that for the electron that is slowed down during the interaction, the radiation cone will become larger, according to the change of the Lorentz factor. 
Moreover, the largest peak in the energy spectrum is generated at the center of the
laser pulse where the intensity is largest and therefore the Lorentz factor takes its minimum value.
Taking this into account, one can write for the normalized photon energy as a function of the scattering angle the following expression:
\begin{align}
y(\vartheta) = \frac{1}{1+\gamma_\star^2\vartheta^2}\,. \label{eq:y_gamma_star}
\end{align}
The spectral lines thus ``straighten out'' for larger values of $a_0$ in the case of compensated nonlinear CS, as compared to the case of linear CS.
This effect can be seen in Fig.~\ref{fig:collimation} (a) and (b), where the energy-angular spectrum of the radiation for the case of $a_0=1$ and $a_0=2$ are presented, respectively.
In these figures, the dashed curves depict the modified angular dependence of the spectral lines
for compensated nonlinear CS, Eq.~\eqref{eq:y_gamma_star}, compared to the case of linear CS (dotted curves).
One can see that the radiation spectrum is roughly bounded by these curves, with the maximum close to the straightened compensated nonlinear line.

As a consequence, the radiation that is collected over a fixed collimation angle $\vartheta_c$ has a smaller bandwidth for higher values of $a_0$. The bandwidth reduction can be clearly seen in Fig.~\ref{fig:collimation} (c), where the radiation spectrum, integrated over all scattering angles up to $\vartheta_c=1/(2\gamma)$ is shown as a function of normalized photon frequency $y$. The reduction of the bandwidth is evident and quantified by the relative FWHM bandwidth $\kappa$ which drops from $\kappa=\unit{20}{\%}$ in the linear CS regime ($a_0=0.01$) down to $\kappa=\unit{2}{\%}$ for the compensated nonlinear CS with $a_0=5$.

The effect of radiation spectrum's angular straightening has an important practical implication. During the design of the x- and gamma-ray photon sources, one specifies the central photon energy as well as the desired FWHM bandwidth $\kappa$. According to \cite{Rykovanov2014}, in the case of the linear CS the approximate optimal collimation angle can be found from the following equation
\begin{align}
\frac{1}{1+\gamma^2 \vartheta_c^2} = 1-\kappa \,. \label{eq:collimation_angle}
\end{align}
Again, due to the fact that the electron slows down during the interaction, the Lorentz factor in the above equation must be replaced by $\gamma_\star$. Thus, the optimal collimation angle for obtaining the bandwidth $\kappa$ in the case of the compensated nonlinear CS increases to $\vartheta_{\star,c}=\vartheta_c\sqrt{1+a_0^2/2}$. 

As one can see, for larger values of $a_0$, it is important to pick the correct, increased collimation angle given by $\vartheta_{\star,c}$.
This provides roughly a factor of $1+a_0^2/2$ more photons in the specified bandwidth compared to the case when the collimation angle is chosen ``incorrectly'' to the value of $\vartheta_c$ according to Eq.~(\ref{eq:collimation_angle}).

\resub{To be specific, we estimate the x-ray photon yield per electron 
to be $N_X\simeq \pi \alpha a_0^2$ in the natural bandwidth
$\kappa = 1/{\omega_0\Delta}$.
This result is consistent with the photon yield of synchrotron radiation \cite{book:Xray,book:Wiedemann},
where $N_X\simeq \pi \alpha K^2/(1+K^2/2)$ photons are
emitted by an electron in an undulator with strength parameter $K$, the undulator's analog of the normalized vector potential $a_0$.
The $1+a_0^2/2$-fold increase in total photon yield is
due to the increased collimation angle in the case of the compensated nonlinear Compton source.
}

\subsection{Effects of electron beam energy spread and emittance}

The results presented in the previous sections were obtained in the approximation of the plane waves and for a single electron. In experiments, however, both electron and laser photon beams always have some energy spread and emittance as well as finite sizes. For the case of the linear CS the effects of electron and laser beam properties on the radiation spectrum were studied in detail, for example, in Refs.~\cite{Sun2011,Rykovanov2014}. In the case of the compensated nonlinear CS, additional considerations are required.

As in the case of the linear CS, the effect of the electron beam energy spread $\sigma_{\gamma,\FWHM}$ on the photon bandwidth is approximately given by
\begin{align}
\Delta y \approx \frac{2\sigma_{\gamma,\FWHM}}{\gamma}\,.
\end{align}
The contribution of the electron beam emittance to the spectral bandwidth differs in the case of compensated nonlinear CS compared to the case of linear CS. As discussed in Sect.~\ref{subsect:monochromatization}, due to the slowing down of the electron, the radiation cone has a wider opening angle. This leads to a more relaxed requirement on the electron beam divergence for obtaining a desired FWHM relative bandwidth $\kappa$.
As shown in Ref.~\cite{Rykovanov2014}, given the required bandwidth $\kappa$,
the electron beams
FWHM angular divergence $\sigma_{\theta,\FWHM}$ has to fulfill the approximate requirement $\gamma^2\sigma_{\theta,\FWHM}^2/4<\kappa$. Replacing $\gamma$ with the interaction modified Lorentz factor $\gamma_\star$, we obtain a relaxed condition for the electron beam divergence in the case of compensated nonlinear CS:
\begin{align}
\frac{\gamma^2\sigma_{\theta,\FWHM}^2}{4\left(1+a_0^2/2\right)}<\kappa\,.\label{eq:divergence_condition}
\end{align}
In the case of laser plasma accelerated electron beams, the electron's angular divergence is one of the main broadening mechanisms for CS photon sources, and its control, i.e. with the help of magnetic or plasma lenses, is required in the source design~\cite{Rykovanov2014}. The compensated nonlinear CS mechanism may help to further reduce the requirement on the electron beam divergence, and it provides additional benefits for compact CS photon sources.

Additional new sources of broadening of the x-ray spectrum emerge in the case of the compensated nonlinear CS.
For instance, electrons traveling under the angle $\theta$ with respect to the beam axis, emit narrow-bandwidth compensated radiation in the direction of their propagation, i.e.~under the angle $\theta$. The on-beam-axis radiation for such electrons (i) has lower photon energy in accordance with Eq.~(\ref{eq:y_gamma_star}) and is taken into account in Eq.~(\ref{eq:divergence_condition}); and
(ii) possesses additional broadening due to the non perfect compensation. For the electron beam with FWHM divergence $\sigma_{\theta,\FWHM}$, the latter effect on the relative spectrum broadening can be estimated by differentiating Eq.~(\ref{eq:compensation.condition}) with respect to $\vartheta_0$, and is on the order of $a_0^2\sigma_{\theta,\FWHM}^2$. For $a_0^2\ll \gamma$, broadening due to this effect is negligibly small compared to Eq.~(\ref{eq:divergence_condition}). 

An additional source of broadening is the finite transverse size
of both electron and laser beams. Unless the laser beam has a rectangular transverse shape, electrons that are traveling off-axis will experience smaller values of $a_0$ and the chirped frequency will be overcompensating the effect from the electron slow-down. Assuming the laser transverse profile to be Gaussian with $a(r)=a_0\exp\left(-r^2/2w_0^2 \right)$ and assuming an electron beam with FWHM transverse size $\sigma_{r, \FWHM}$, additional relative spectral broadening will be approximately given by
\begin{align}
\Delta y & = \frac{1+\frac{1}{2}a_0^2}{1+\frac{1}{2}a_0^2 e^{\left(-\sigma_{r,\FWHM}^2/w_0^2 \right)}}-1 \nonumber \\
& \approx e^{\left(\sigma_{r,\FWHM}^2/w_0^2 \right)} - 1 \,,
%\approx \frac{\sigma_{r,\FWHM}^2}{w_0^2}\,,
\end{align}
where the latter approximation works well for $a_0^2\gg 1$. One can see that non-perfect frequency compensation due to focusing can lead to considerable additional broadening and must be considered when designing a source. 

Moreover, additional broadening arises due to {pulse diffraction}:
If the pulse's temporal shape is Gaussian in the focus and one compensates the frequency according to this shape, the electron will see a distorted temporal shape while moving through the interaction region due to the laser pulse diffraction. However, we expect this effect to be small for $w_0\ll \lambda_L$, where $w_0$ is the laser spot size and $\lambda_L$ is the laser wavelength. Moreover, for higher photon yield it is beneficial to use waveguides~\cite{Rykovanov2014}, where diffraction is avoided.

Estimations of the effects of the electron and laser beam parameters on the spectrum broadening show that compensated nonlinear CS is feasible to be used in experiments to achieve high photon yields. To check these predictions we have performed numerical simulations using the code VDSR~\cite{Chen2013} with realistic electron and laser beam parameters:
The electron beam with the central Lorentz factor $\gamma=529$ has a $\unit{2.2}{\%}$ energy spread and
normalized emittance $\varepsilon_n\approx \unit{0.2}{\milli\metre\usk\milli\radian}$,
i.e.~its transverse radius is $\sigma_{r,\FWHM} =  \unit{1.8}{\micro \metre}$,
its duration is $\unit{10}{\femto\second}$ and
the angular divergence is $\sigma_{\theta,\FWHM}=\unit{0.2}{\milli\radian}$. These parameters are realistic for the current laser plasma accelerated electrons~\cite{Esarey2009, Plateau2012, Weingartner2012, Lundh2011}, which show great promise towards compact photon sources.
The laser beam has a Gaussian shape in every direction, normalized peak amplitude $a_0=2.83$,
and r.m.s.~radius $w_0=\unit{30}{\micro\metre}$.
The asymptotic wavelength was chosen to be $\lambda_L=\unit{0.8}{\micro\metre}$ ($\omega_0=\unit{1.55}{\electronvolt}$).
The laser pulse's frequency is properly chirped according to
Eq.~(\ref{eq:optimal.theta0}) for on-axis compensation, i.e.~$\vartheta_0=0$. We have also assumed that the laser pulse propagates in a waveguide, thus is not divergent. In order to create a proper model for the electromagnetic fields, including diffraction, one would need to take into account that every color of the laser pulse has a different Rayleigh range and this is outside the scope of this manuscript. According to the estimations provided earlier in this section, the total broadening due to electron beam energy spread, divergence and focusing should be on the order of $\unit{5}{\%}$. 

\begin{figure}[!ht]
\includegraphics[width=0.99\columnwidth]{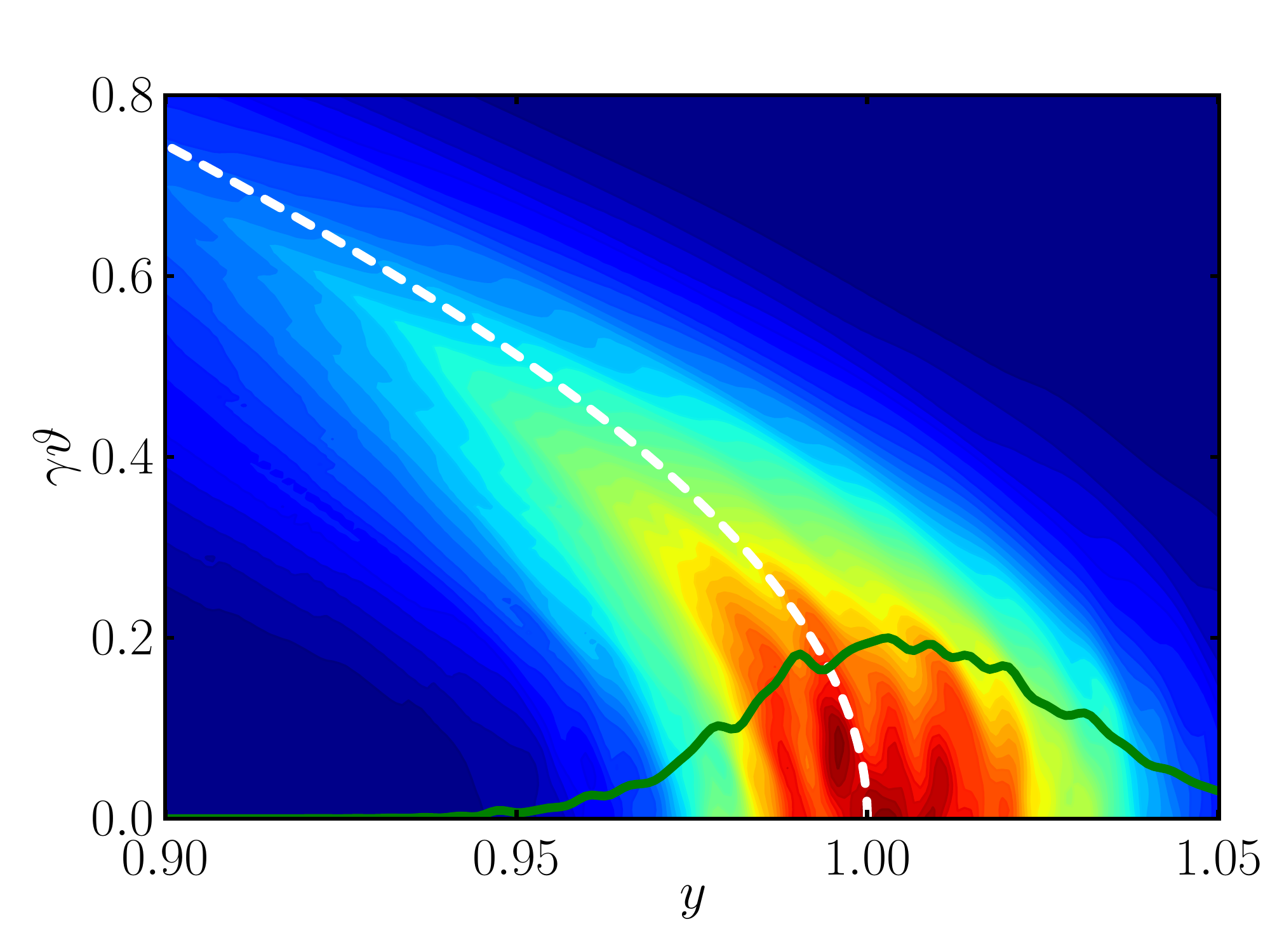}
\caption{Simulated energy and angular radiation spectrum of a realistic electron beam interacting with a focused laser pulse with peak intensity $a_0=2.83$. The white dashed curve shows the angular dependence of the radiation's energy according to Eq.~(\ref{eq:y_gamma_star}). The solid green line depicts the on-axis line-out of the radiation spectrum.
}
\label{fig:vdsr_simulation}
\end{figure}

The results of the simulations are presented in Fig.~\ref{fig:vdsr_simulation} where the energy-angular radiation spectrum is shown (in arbitrary units). The white dashed line shows the curve given by Eq.~(\ref{eq:y_gamma_star}) for $a_0=2.83$ and fits well with the simulations.
The green solid line outlines the on-axis spectrum.
One can see that the spectrum is narrow and has an on-axis bandwidth of approximately $\unit{5}{\%}$ in good agreement with the estimates provided above. The main contribution to the bandwidth is in this case due to the electron energy spread and is approximately $\unit{4.4}{\%}$. The contributions due to electron beam divergence and focusing are approximately $\unit{0.5}{\%}$ each.
It is worth mentioning that without the compensation the ponderomotive broadening due to the nonlinear
effect would be on the order of $\unit{80}{\%}$ \cite{Rykovanov2014}.
This proof-of-principle simulation takes into account realistic electron and laser beams and demonstrates that compensated nonlinear CS is a promising route towards intensifying the total photon yield of x- and gamma-ray sources.

\section{Conclusions}

\label{sect:summary}

In this paper, we analyzed the ponderomotive broadening of the
spectral lines in nonlinear Compton scattering of intense laser light on ultra-relativistic electrons,
and how this effect can be compensated by a suitably chirped initial laser pulse.
In our analysis, we fully take into account the effects of electron recoil and
spin by working within the framework of strong field quantum electrodynamics
in the Furry picture.
We systematically derive a prescription for the optimal frequency modulation that
allows to compensate the ponderomotive broadening effect.
We find that this optimal frequency modulation of the initial laser pulse, calculated within the framework of relativistic quantum electrodynamics, coincides with
the result found previously in the literature in the recoil-free Thomson limit.
Our analysis shows that a complete compensation of the ponderomotive broadening is possible just for one single
scattering angle, for both the fundamental line and all higher harmonics.
Thus, over a finite angular region one can achieve at most a partial compensation of the ponderomotive
broadening.

Because the electron becomes slower as it enters the high-intensity laser pules, the partly compensated
nonlinear Compton spectral peaks follow a
modified frequency-angle correlation,
which makes the frequency of the scattered photons less sensitive on the scattering angle.
As a consequence the emitted 
radiation in compensated nonlinear Compton scattering
becomes more monochromatic over a larger angular region,
which allows for a larger collimation angle for the scattered x-rays
and provides more photons in a given bandwidth.

Because the optimal frequency modulation is derived using the idealized case of a single electron colliding with a plane-wave laser we also
study the influence of realistic electron and laser beams on the compensation of the ponderomotive
broadening.
In particular, we found that the impact of the electron beam angular divergence on the bandwidth
of the x-ray source is even reduced due to the modified frequency-angle correlation.
Our analysis
shows that the compensation of ponderomotive broadening by chirped laser pulses
is a promising route towards operating narrowband Compton scattering x- and gamma-ray sources
at high laser intensity.

\begin{acknowledgments}
This research was partially conducted during the KITP program ``Frontiers of Intense Laser Physics''.
During that period D.S. was supported in part by the National Science Foundation under Grant No. NSF PHY11-25915. S.G.R. is supported by the Helmholtz Gemeinschaft (Nachwuchsgruppe VH-NG-1037).
\end{acknowledgments}

\appendix

\section{Choice of Initial Conditions for the Optimal Frequency Modulation}

In Section \ref{sect:compensation} we found the optimal frequency modulation
\eqref{eq:compensation.solution}
as solution of a differential equation
\eqref{eq:compensation.dgl}, that
relates the laser pulse envelope $g(x^+)$ to the laser chirping $\omega(x^+)$.
In order to get a unique solution we need to provide some initial conditions for the solution of the differential equation.
In this Appendix we discuss our choice in comparison to those used in the Ref.~\cite{Terzic:PRL2014}.

\begin{figure}[!th]
\vspace*{1ex}
\includegraphics[width=0.99\columnwidth]{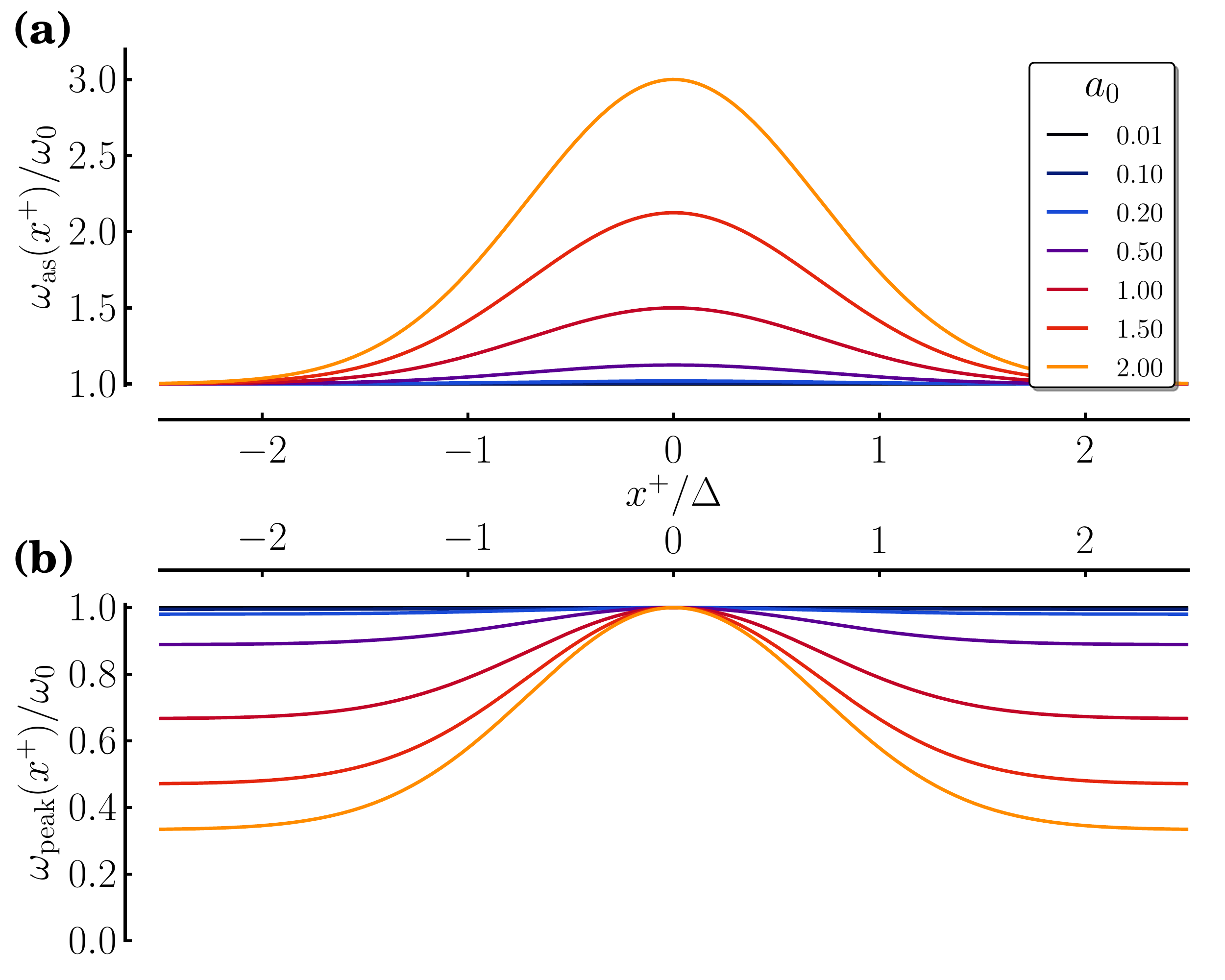}
\caption{(Color online)
Comparison of the functional dependence of the optimal frequency modulation $\omega(x^+)$
as a solution of the differential
equation for various
values of $a_0$ for \eqref{eq:compensation.dgl} for asymptotic initial conditions (a) and peak initial conditions (b).
}
\label{fig:ic_comparison}
\end{figure}

In this paper we employ the initial conditions that are fixed at asymptotic times $x^+_i\to -\infty$ where the ponderomotive four-potential vanishes, $U^\mu(-\infty)=0$, as $\omega(-\infty) = \omega_0$.
With this choice, the frequency modulation reads
\begin{align}
\omega_{\rm as}(x^+) = \omega_0 \left( 1 + \beta_0 g^2(x^+) \right) \,,
\label{eq:optimal.chirp.asymptotic.app}
\end{align}
and the chirped frequency is always larger than $\omega_0$, see Fig.~\ref{fig:ic_comparison} (a).
This choice of initial conditions allows to easily compare the case of the compensated nonlinear
CS with the case of linear linear CS, since
the compensated lines condense at the linear Compton lines at $\omega'=\ell \Omega/(1+\ell \chi)$
for any value of $a_0$, and where $\Omega$, and $\chi$ are defined below Eq.~\eqref{eq:omega.SPA.0}, and $\beta_0$ in Eq.~\eqref{eq:def.beta0}.

A different choice of initial conditions, which has been
employed in Ref.~\cite{Terzic:PRL2014},
is fixed at the peak of the laser pulse $x^+_i=0$ as $\omega(0) = \omega_0$.
Thus, the optimal frequency modulation reads
\begin{align}
\omega_{\rm peak}(x^+) = \omega_0 \frac{1+\beta_0 g^2(x^+)}{1+\beta_0} \,,
\label{eq:optimal.chirp.center}
\end{align}
and the chirped frequency is always smaller than $\omega_0$, see Fig.~\ref{fig:ic_comparison} (b).
In particular, when optimizing for the on-axis radiation we get
\begin{align*}
\omega_{\rm peak}(x^+) = \omega_0 \frac{1+\frac{a_0^2}{2}g^2(x^+)}{1+\frac{a_0^2}{2}} \,,
\end{align*}
which exactly coincides with \cite{Terzic:PRL2014}.
We mention here again that our derivation of the optimal frequency modulation takes into
the electron recoil exactly
by working within the framework of relativistic strong-field quantum electrodynamics,
while the recoil is neglected in the classical treatment of \cite{Terzic:PRL2014}.
For this choice of initial conditions the compensated lines condense at the maximally red-shifted nonlinear Compton lines at $\omega'=\ell \Omega/(1+\ell \chi + \beta_0)$.
It is, thus, more complicated to compare with the case of linear Compton scattering as
the position of the compensated spectral lines changes with $a_0$ and a comparison
with linear CS would require an $a_0$-dependent redefinition of $\omega_0$ for each case.

%\bibliographystyle{../../revtex/apsrev4-1}
%\bibliography{library}

\end{document}